\documentclass{iopjournal}

\usepackage{amsmath, amssymb, bm, orcidlink, ragged2e}

\usepackage[numbers,sort&compress]{natbib}
\begin{document}

\justifying

\articletype{Paper}

\title{Spinodal-like scaling behavior after a temperature quench 
  across the first-order phase transition in three-dimensional
  $q$-state Potts models}

\author{Andrea Pelissetto$^{1,*}$\orcidlink{0000-0002-3633-0496},
  Davide Rossini$^{2,*}$\orcidlink{0000-0002-9222-1913}
  and Ettore Vicari$^{3,*}$\orcidlink{0000-0002-7469-9614}}

\affil{$^1$Dipartimento di Fisica dell'Universit\`a di Roma
  La Sapienza and INFN, Sezione di Roma I, I-00185 Roma, Italy}

\affil{$^2$Dipartimento di Fisica dell'Universit\`a di Pisa
  and INFN, Largo Pontecorvo 3, I-56127 Pisa, Italy}

\affil{$^3$Dipartimento di Fisica dell'Universit\`a di Pisa,
  Largo Pontecorvo 3, I-56127 Pisa, Italy}

\affil{$^*$Authors are listed in alphabetic order.}

\email{davide.rossini@unipi.it}

\keywords{Classical phase transitions, Dynamical processes, Nucleation}

\begin{abstract}
  \justifying
  We study the out-of-equilibrium spinodal-like behavior of
  three-dimensional (3D) $q$-state Potts models (for $q\ge 3$),
  observed when the temperature is quenched across the first-order 
  transition (FOT) point $\beta_{\rm fo}=T_{\rm fo}^{-1}$.
  We consider a standard quench protocol, in which high-temperature
  configurations, thermalized at $\beta_i<\beta_{\rm fo}$, are driven
  across the FOT by a purely relaxational dynamics at $\beta>\beta_{\rm fo}$.
  We focus on the emergence of spinodal-like behaviors in the thermodynamic
  limit, associated with the dynamic phase change.  We argue that, if
  the nucleation of smooth droplets is the relevant mechanism of the
  post-quench phase change, for sufficiently small $\beta_{\rm fo}-\beta_i>0$, 
  the time-dependent energy density should scale in terms of
  $\rho = (\ln t)^{3/2} \delta$, where  $\delta = \beta/\beta_{\rm fo}-1$,
  with a discontinuity at a particular value $\rho=\rho_s>0$.
  This implies the emergence of a spinodal-like behavior, whose time
  scale $\tau$ increases exponentially as
  $\ln \tau \approx (\rho_s/\delta)^{2/3}$ in the limit $\delta\to 0^+$.
  We present a numerical analysis of the quench protocol 
  in the 3D $q=6$ Potts model, which supports the above
  spinodal-like scenario.
\end{abstract}

\section{Introduction}
\label{intro}

Many-body systems exhibit distinct out-of-equilibrium 
dynamic behaviors  when driven across first-order transitions (FOTs); 
these include hysteresis, coarsening, and the emergence of 
dynamic spinodal-like phase changes (see, e.g., 
Refs.~\cite{Binder-87, Bray-94, CA-99, RV-21, PV-24, RTMS-94, MM-00,
  LFGC-09,NIW-11,EH-13, PV-15, PV-16, PV-17, PV-17b, LZ-17, PPV-18,
  SW-18, PRV-18, Bar-etal-18, PRV-18-def, Fontana-19, LZW-19,
  PRV-20, DRV-20, CCP-21, SCD-21, CCEMP-22, TV-22, TS-23,
  PRV-25, PV-26, PRV-26, PRV-26-2, PRV-26-3}).
In this paper, we consider the three-dimensional (3D) $q$-state
Potts model and discuss its out-of-equilibrium behavior
when the temperature is instantaneously quenched across
the thermal FOT present for $q\ge 3$~\cite{Wu-82}.
For this purpose, we consider a standard quench protocol. We start at
$t=0$ from an ensemble of configurations thermalized at inverse 
temperature  $\beta_i < \beta_{\rm fo}$, where $\beta_{\rm fo}$ is 
the inverse temperature of the first-order transition. The system
then evolves under a purely relaxational dynamics at fixed 
$\beta>\beta_{\rm fo}$, eventually ordering as time increases. 
We observe the emergence of a spinodal-like behavior in 
the thermodynamic limit, quite similar to that observed 
when the temperature is instead slowly varied across the FOT~\cite{PRV-26}. 
We identify the relevant timescale of the post-quench evolution and 
show that it can be derived by assuming that droplet nucleation
is the relevant mechanism responsible for the phase change.

Droplet nucleation is apparently the relevant mechanism that explains
the out-of-equilibrium dynamics when crossing a FOT in several
different systems. This the case in 2D Ising systems, when the
magnetic low-temperature FOT line is crossed, either very slowly or
instantaneously~\cite{PV-26,PRV-26-3}, and also in 2D and 3D $q$-state
Potts models, when slowly crossing their thermal
FOTs~\cite{PV-17,PRV-26}. However, the scaling behaviors observed when
crossing the low-temperature FOT line in 3D and 4D Ising
systems~\cite{PV-26} are inconsistent with the simplest droplet
nucleation scenario, suggesting that different mechanisms may be at
work, somehow depending on the nature of the FOT. The study reported
in this paper is meant to further investigate this issue and add
information on the out-of-equilibrium spinodal-like phenomena at FOTs.

Assuming that the slowest timescale of the post-quench phase change is
provided by the nucleation of smooth droplets, as verified in the case
of slow-crossing protocols~\cite{PRV-26}, we argue that the
post-quench time dependence of the energy density in the thermodynamic
limit satisfies the scaling behavior $E(t,\beta) \approx {\cal E}(\rho)$
in terms of the variable $\rho \sim (\ln t)^\kappa \delta$
where $\delta=\beta/\beta_{\rm fo}-1$ and $\kappa=3/2$ (for
sufficiently small differences $\beta_{\rm fo}-\beta_i$).  Moreover,
the scaling function ${\cal E}(\rho)$ exhibits a singularity at a
particular value $\rho=\rho_s>0$ (corresponding to $\delta>0$), where
the energy scaling function ${\cal E}(\rho)$ is discontinuous in the 
thermodynamic limit, varying between the equilibrium values 
of the energy at the FOT.  This implies that the timescale
$\tau$ of the passage from the disordered to the ordered phase
increases exponentially as $\tau \approx \exp[(\rho_s/\delta)^{2/3}]$
in the limit $\delta\to 0^+$. To verify these predictions for 
the post-quench dynamics of Potts systems, we consider the 3D $q=6$ Potts
model and study its evolution under a purely relaxational dynamics---we 
use a standard heat-bath Monte Carlo (MC) algorithm.
The results nicely confirm the above predictions, thus confirming
that the nucleation of smooth droplets is the relevant mechanism for
the out-of-equilibrium phase change, when the temperature is suddenly 
quenched across the FOT.

The paper is organized as follows. In Sec.~\ref{moddyn} we introduce
the 3D $q$-state Potts model and the quench protocol.
In Sec.~\ref{TLsca} we derive some quantitative scaling predictions
under the hypothesis that the post-quench dynamic phase
change is driven by the nucleation of smooth droplets, providing
the largest time scale of the process. In Sec.~\ref{numres} we
support these spinodal-like scaling predictions by presenting a
numerical MC analysis of the quench dynamics across the FOT of the 3D
$q=6$ Potts model. Finally, in Sec.~\ref{conclu} we summarize and draw
our conclusions.

\section{Model and quench protocol}
\label{moddyn}

\subsection{The 3D $q$-state Potts model}
\label{model}

The Hamiltonian and partition function of the 3D nearest-neighbor
$q$-state Potts model are~\cite{Wu-82}
\begin{equation}
  H = J \sum_{\langle {\bm x}{\bm y}\rangle}
  \left[1-\delta(s_{{\bm x}}, s_{ {\bm y}})\right] , \qquad
  Z = \sum_{\{s_{\bm x}\}} e^{-\beta H} , 
  \label{potts}
\end{equation}
where the sum in $H$ is over the nearest-neighbor sites
$\langle {\bm x}{\bm y}\rangle$ of a cubic lattice, $s_{\bm x}$ are
integer variables $s_{{\bm x}}=1, 2, \ldots, q$, $\delta(a,b)=1$ if
$a=b$ and zero otherwise, while $\beta=1/T$.
We consider systems of size $L$ in each direction with periodic
boundary conditions, which preserve the $q$-state permutation symmetry.
Moreover, we set $J=1$ without loss of generality.

The 3D Potts model undergoes a FOT at a finite $\beta=\beta_{\rm fo}$
for any $q\ge 3$, between a disordered phase and $q$ equivalent
ordered phases~\cite{Wu-82, BBD-08}. This FOT is characterized by
a discontinuity in the energy density and magnetization.
We focus on the rescaled energy density
\begin{equation}
  E(\beta)\equiv {q\over 3(q-1)} {\langle H \rangle\over V} ,
  \qquad V = L^3,
  \label{ener}
\end{equation}
which varies  between $E=1$ (for $\beta\to 0$) and  $E=0$ (for
$\beta\to\infty$). In the thermodynamic limit, $E(\beta)$ is
discontinuous: 
\begin{equation}
  {\rm lim}_{\beta\to\beta_{\rm fo}^{\pm}}  \; {\rm lim}_{V\to\infty} \;
  E(\beta) = E_{\pm}, 
  \label{deltae}
\end{equation}
with a rescaled latent heat $\Delta_h \equiv E_+ - E_- > 0$.
Thus, energy-density values $E>E_+$ correspond to a system in the
high-temperature disordered phase, and $E<E_-$ to the low-temperature
ordered phase. In the absence of perturbations that break the
$q$-state permutation symmetry, the magnetization vanishes for any
$\beta$ (the observation of its discontinuity requires a particular
limit involving an external magnetic field).  Therefore, for periodic
boundary conditions and in the absence of symmetry-breaking
perturbations, systems in the low-temperature phase develop large
ordered regions associated with different $q$ states, to ensure a
vanishing magnetization.

\subsection{Quench protocol across the transition}
\label{quenchprot}

In our study we consider a standard quench protocol. The evolutions
start at $t=0$ from ensembles of high-temperature configurations
thermalized at $\beta_i<\beta_{\rm fo}$ in the disordered phase. In
particular, we choose $\beta_i/\beta_{\rm fo} = 1 - \epsilon$ with
$0<\epsilon\ll 1$, thus effectively $\beta_i=\beta_{\rm fo}^-$ in the
thermodynamic limit (other values of $\epsilon$ within
$0<\epsilon\lesssim 10^{-3}$ do not lead to significant quantitative
changes in the post-quench evolution in the infinite-volume
limit). Then, for $t>0$, the system evolves under a purely
relaxational dynamics at fixed $\beta >\beta_{\rm fo}$, corresponding
to the low-temperature phase. This can be considered as an
instantaneous temperature quench across the thermal FOT.

We consider a heat-bath dynamics~\cite{Binder-76, Creutz-book} (at each
site, a new spin is chosen using the conditional probability
distribution at fixed neighboring spins), which is a specific example
of a purely relaxational dynamics.  Spins are updated using a
checkerboard scheme. Since cubic-like lattices are bipartite, sites
can be divided in two sets, the set of even and the set of odd sites,
respectively. We first update all spins at even sites, then all spins
at odd sites.  A time unit corresponds to a complete lattice sweep.

We focus on the quench dynamics in the thermodynamic limit, by
determining the time evolution in the infinite-size limit keeping
$\beta$ fixed, then we study the scaling behavior in the small
$\delta$ limit.  Information on the dynamics is provided by the
fixed-time average rescaled energy density
\begin{equation}
  E(t,\beta,L)= {q\over 3(q-1)} {\langle H \rangle_t\over V}.
  \label{etts}
\end{equation}
The average $\langle H \rangle_t$ at time $t$ is performed over
a large number of stochastic trajectories, starting from thermalized
configurations at inverse temperature $\beta_i$ and arising
from the heat-bath relaxational dynamics.
In practice, we consider one trajectory for each initial
equilibrium configuration, so that all trajectories we analyze
are independent if the initial configurations are decorrelated.

\section{Spinodal-like scaling  behavior in the thermodynamic limit}
\label{TLsca}

In this section we argue that the post-quench relaxational dynamics
across the FOT of Potts models develops a dynamic spinodal-like scaling
behavior in the thermodynamic limit, which can be understood 
by assuming that droplet nucleation is the relevant dynamic
mechanism.

Before presenting the argument, we recall that in the mean-field
theory of FOTs~\cite{Binder-87}, the
free-energy density near the transition is characterized by two
distinct minima. They represent a stable and a metastable state,
separated by a free-energy barrier that diverges as $L^d$ in the
thermodynamic limit. From a dynamical perspective, this implies that
the system can remain trapped in the metastable state for a duration
that diverges as $L \to \infty$. Within this framework, the spinodal
line is defined as the boundary where the metastable state becomes
unstable, i.e., the point at which the local free-energy minimum
corresponding to the metastable state disappears. In short-range
models, there are no thermodynamic metastable states for any
$\beta>\beta_{\rm fo}$~\cite{Binder-87}. Nevertheless,
metastability is a ubiquitous feature of the dynamics of systems
close to FOTs: for instance, it emerges naturally
when the evolution drives the system across a FOT,
as is the case in this work.  This is due to the fact
that equilibration requires the system to go across regions of
atypical configurations, i.e., to overcome free-energy
barriers. These barriers are finite even in the infinite-volume
limit---this is at variance with the metastable states defined in
the mean-field approach. Consequently, the dynamic metastable state
decays in a finite time even as $L\to\infty$. In the following, we
present an argument that will allow us to determine the typical time
scale of this process.
    
Let us assume that the transition to the ordered phase starts with the
nucleation of droplets of the $q$ ordered phases, of size $R_{\rm dr}
\ll L$ for a large system size $L$, and that the typical droplet
radius is larger than the correlation length of the local
fluctuations.  The volume of each droplet is of order $R_{\rm dr}^d$
(where $d$ is the spatial dimension of the system), while the area
scales as $R_{\rm dr}^{d-1}$ if they have a smooth boundary.  The time
$t$ needed to create a droplet of size $R_{\rm dr}$ should increase
exponentially with its area~\cite{Binder-87, RTMS-94, PV-17}, so $\ln
t \sim R_{\rm dr}^{d-1}$, leading to the relation $R_{\rm dr}(t) \sim
(\ln t)^{1/(d-1)}$.  Since all spins in the droplet have the same
color, the energy of the droplet is proportional to its volume $R_{\rm
  dr}^d$. Assuming that the nucleation of such smooth droplets
represents the relevant mechanism for the post-quench dynamics, the
basic scaling variable of the time evolution is expected to be related
to the ratio between the ordering energy and the thermal energy, i.e.,
\begin{equation}
  R_{\rm dr}^d(t) \, \delta,\qquad
  \delta = {\beta-\beta_{\rm fo}\over \beta_{\rm fo}}.
  \label{dtprod}
\end{equation}
Therefore, the relevant time scaling variable is expected to be
\begin{equation}
  \rho = (\ln t)^\kappa\delta,
  \qquad \kappa = {d\over d-1}={3\over 2}.
  \label{rhodef}
\end{equation}
Under the hypothesis that the nucleation mechanism of smooth droplets
is the one providing the largest timescale, the post-quench evolution
of the energy density is thus expected to scale as
\begin{equation}
  E(t,\beta)\approx {\cal E}(\rho).
  \label{etsca}
\end{equation}
If Eq.~(\ref{etsca}) holds, the phase flip occurs at a specific
value of $\rho$. For instance, we can define the timescale $\tau$ as
the time where $E(\tau,\beta) = (E_+ + E_-)/2$. Then,
Eq.~(\ref{etsca}) implies that $\tau$ depends on $\delta$ in such a
way to keep $\rho_s = (\ln \tau)^\kappa\delta$ constant, i.e., it
implies
\begin{equation}
  \tau \approx \exp[(\rho_s/\delta)^{2/3}]
  \label{taubeh}
\end{equation}
for $\delta \to 0^+$.

As already observed when considering a slow dynamics across the
FOTs of 2D Ising and Potts
models~\cite{PV-26,PRV-26-3,PV-17,PRV-26}, the scaling function
${\cal E}(\rho)$ may be discontinuous.  This occurs if the phase
change is driven by a different and much faster mechanism, i.e., if
the typical time $\tau_{pc}$ needed by the system to go from one
phase to the other phase satisfies $\tau_{pc}/\tau\to 0$ for
$\delta\to 0^+$.  In this case, the scaling function ${\cal E}(\rho)$
is discontinuous at $\rho=\rho_s$. Indeed, if $E(t) = E_+$ for
$t < t_- = \tau - \tau_{pc}$ and $E(t) = E_-$ for
$t < t_+ = \tau + \tau_{pc}$, we have
$\rho_\pm = \delta (\ln t_\pm)^\kappa \to \rho_s$ for $\delta \to 0$:
the phase change is essentially
instantaneous in the variable $\rho$. Moreover, if we additionally
assume a power-law suppression for the ratio $\tau_{pc}/\tau$, i.e.,
${\tau_{pc}/\tau}\sim \delta^{\omega}$ for $\delta \to 0$, then we
have $\rho_\pm - \rho_s \sim \delta^\theta$ with
$\theta=\omega+\kappa^{-1}$, which suggests a scaling behavior in
terms of $\rho_r=(\rho - \rho_s) \delta^{-\theta}$ in the narrow
region in which the phase change occurs. As discussed below,
numerical data for the 3D $q=6$ Potts model nicely support this
picture.

In analogy with the results obtained for the quench dynamics in
2D Ising systems~\cite{PRV-26-3}, we expect the phase change to be
due to the coalescence of the droplets of the stable phase and not
to the independent growth of isolated domains.  Droplets increase
their size by merging, a process that is always energetically
favored. In this case, the very fast passage from the disordered to
the ordered phase should be associated with a percolation transition
of the droplets of the stable phase---preliminary results support
this claim also for the quench dynamics in the Potts model.  This
interpretation of the out-of-equilibrium dynamics has already been
proposed and checked numerically for the Kibble-Zurek dynamics in 2D
Potts and Ising models~\cite{PV-17,PV-26,PRV-26} and, for the quench
dynamics, in 2D Ising systems~\cite{PRV-26-3}.  As we shall show,
it also applies to the 3D Potts model.

\section{Numerical results}
\label{numres}

\begin{figure}[!t]
  \begin{centering}
    \includegraphics[width=0.6\columnwidth, clip]{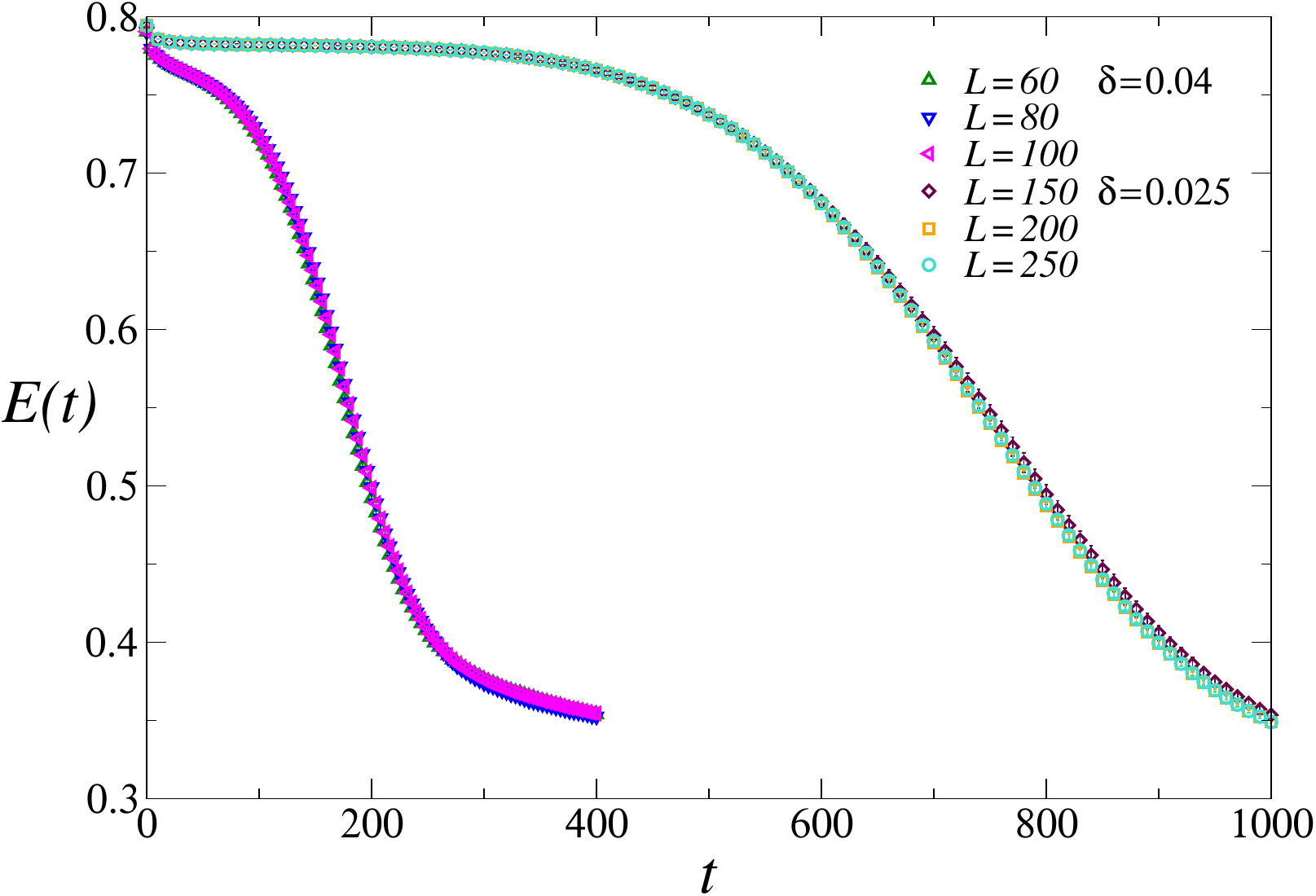}
    \caption{Post-quench time evolution of the rescaled energy density
      $E(t)$ for the 3D $q=6$ Potts model, versus the time $t$ for
      several values of $\delta$ and $L$.  The data for different
      system sizes at fixed $\delta$ converge towards a
      common asymptotic curve, becoming nearly indistinguishable
      on the scale of the figure.
      This curve provides an accurate approximation of the time
      evolution in the thermodynamic limit.
      Statistical errors are barely visible in the figure.}
    \label{rawdata}
  \end{centering}
\end{figure}

We now present a numerical analysis of the quench protocol for the
representative case of the 3D $q=6$ Potts model.
This system undergoes a FOT at the inverse temperature
$\beta_{\rm fo} = 0.739214(3)$, where the rescaled energy density $E$,
defined in Eq.~(\ref{ener}) is discontinuous in the thermodynamic
limit, and the rescaled latent heat takes the value
$\Delta_h = E_+-E_- \approx 0.4729$ (see Ref.~\cite{BBD-08}).
We analyze the post-quench out-of-equilibrium behavior
of the heat-bath dynamics in
the thermodynamic limit, defined as the limit $L\to \infty$ while
keeping $\beta$ fixed, for some values of $\beta$ corresponding to
$\delta=\beta/\beta_{\rm fo}-1$ in the interval $[0.025,0.05]$.  To
this end, we perform MC simulations at fixed $\delta$ for several
system sizes, increasing $L$ until the average energy curves become
independent of $L$ (within the accuracy of the numerical results).  We
only show results for the rescaled energy density $E(t)$,
cf. Eq.~(\ref{etts}), as it provides the most relevant information for
the passage from the high-temperature to the low-temperature
phase. The post-quench evolution of the energy density is reported in
Fig.~\ref{rawdata} for two values of $\delta$ and several values of
$L$, to show the approach to the thermodynamic limit at fixed
$\delta$.  We note that the lattice size necessary to approach the
infinite-volume limit rapidly increases (likely exponentially) with
decreasing $\delta$. For example the curves for $\delta=0.04$ appear
to converge for $L\gtrsim 50$, while they only converge for
$L\gtrsim 200$ when $\delta=0.025$ (within statistical errors).

\begin{figure}[!t]
  \begin{centering}
    \includegraphics[width=0.6\columnwidth, clip]{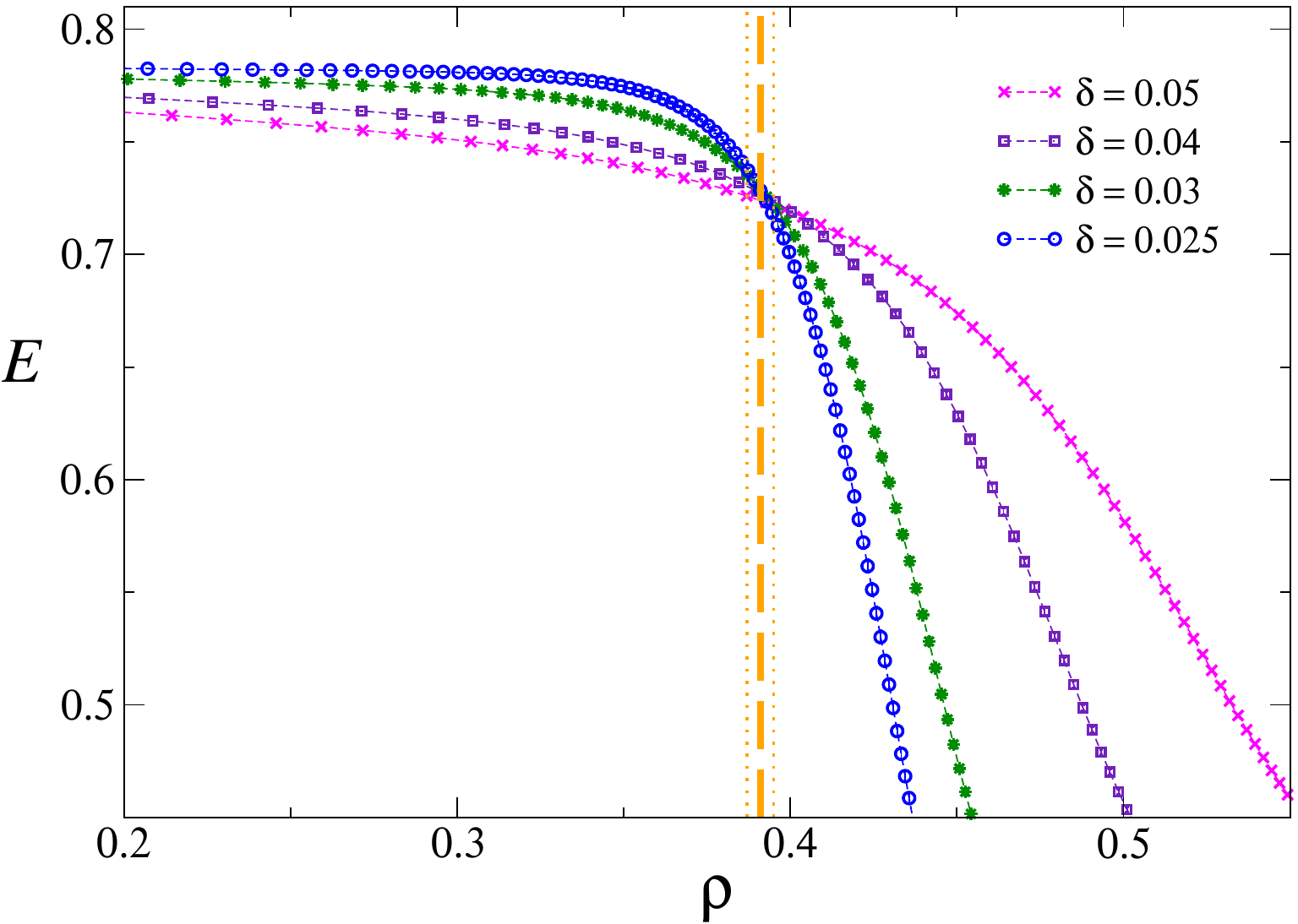}
    \caption{Post-quench evolution of the $q=6$ energy density $E(t)$
      in the thermodynamic limit versus $\rho= (\ln t)^{3/2} \delta$.
      The vertical dashed line corresponds to the estimate
      $\rho_s=0.391(4)$ of the asymptotic crossing point (the interval
      between the dotted lines gives the uncertainty).}
    \label{firstresc}
  \end{centering}
\end{figure}

To check the scaling behavior predicted by the droplet nucleation
mechanism outlined in Sec.~\ref{TLsca}, in Fig.~\ref{firstresc}
we plot the infinite-volume energy density $E(t)$ versus
the scaling variable $\rho$ defined in Eq.~(\ref{rhodef}).
The data are consistent with a singular scaling behavior.
Indeed, all curves for $\delta\in [0.025,0.05]$ have a quite stable
crossing point at an approximately constant
$\rho = \rho_s \approx 0.39$, and become increasingly steeper
close to this point, as $\delta$ decreases. We estimate the crossing
point to be at $\rho_s=0.391(4)$.

\begin{figure}[!t]
  \begin{centering}
    \includegraphics[width=0.6\columnwidth, clip]{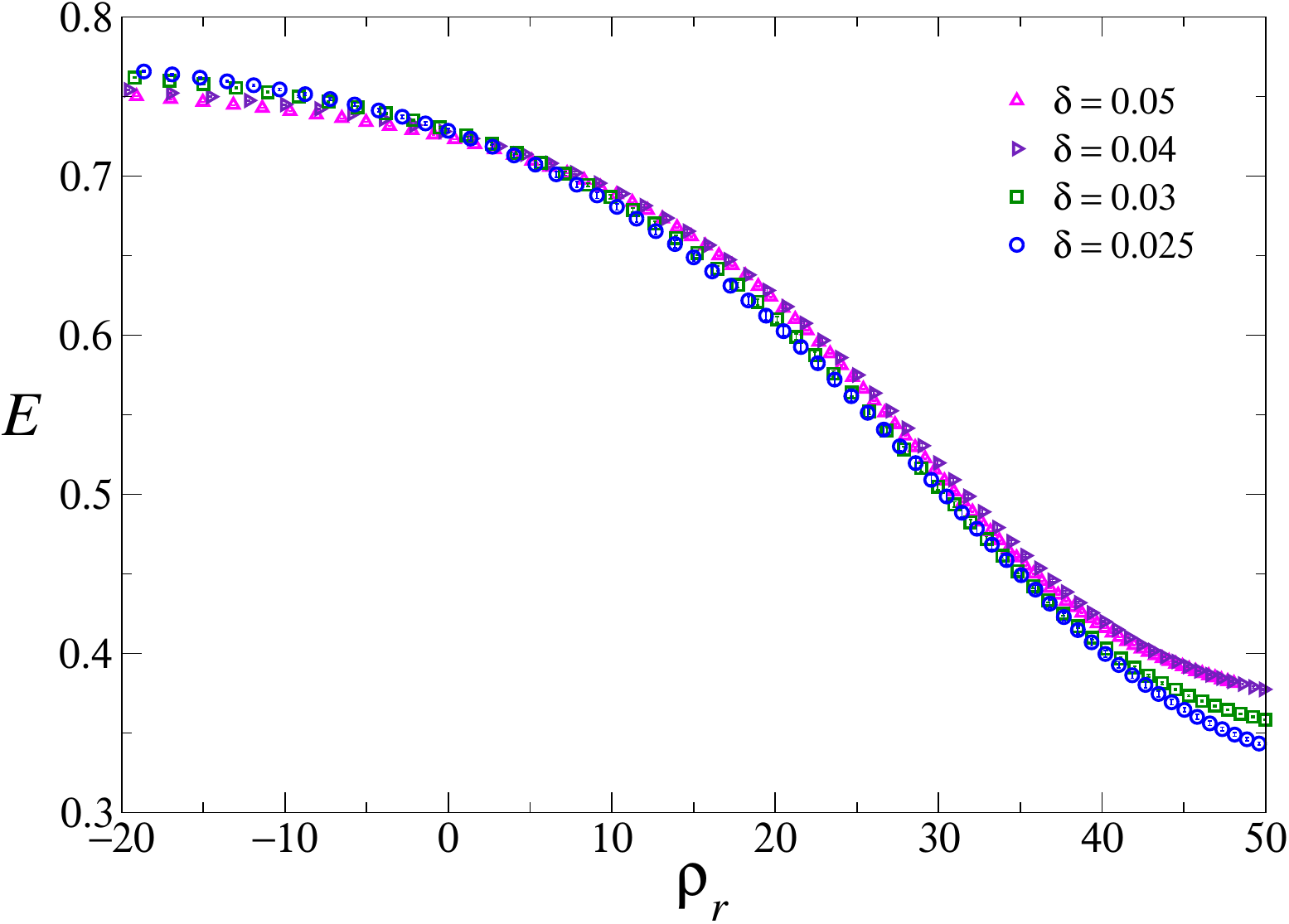}
    \caption{The post-quench energy density $E(t)$ versus
      $\rho_r \equiv (\rho - \rho_s)\,\delta^{-\theta}$, using the optimal
      values $\rho_s=0.391$ and $\theta=1.8$.  The data appear to
      approach an asymptotic scaling curve with decreasing $\delta$.}
    \label{secondresc}
    \end{centering}
\end{figure}

This behavior is consistent with the arguments of Sec.~\ref{TLsca}: it
reflects the fact the transition between the high-temperature and
low-temperature phases is much faster than the typical time required
to nucleate the droplets. As discussed in Sec.~\ref{TLsca},
the presence of a stable crossing point
suggests the emergence of an additional scaling behavior of the energy
density versus
\begin{equation}
  \rho_r = (\rho-\rho_s) \, \delta^{-\theta},
  \label{barsigma}
\end{equation}
where $\theta$ is a new exponent.  This scaling behavior should
  hold in the transition region $\rho \approx \rho_s$, where the
  system changes phase.  This is confirmed by the data shown in
Fig.~\ref{secondresc}: the different curves approximately collapse
onto a single one in a reasonably large interval around $\rho_s$, for
$\theta\approx 1.8$ (data do not allow us to obtain an accurate
estimate of $\theta$; apparently smaller values of $\delta$ are
required to observe the asymptotic behavior).  The scaling in terms of
$\rho_r$ implies that the scaling function ${\cal E}(\rho)$ associated
with the energy density develops an asymptotic spinodal-like
discontinuity at $\rho_s$, with corrections that decay as
$\delta^{\theta}$, as $\delta$ decreases.

We finally mention that a spinodal-like scaling behavior is also
observed for Kibble-Zurek protocols~\cite{PRV-26}, in which the
temperature is slowly varied across the FOT. Also in this case, the
observed behavior is consistent with the predictions obtained by
assuming that the nucleation of smooth droplets is the mechanism
providing the timescale of the phase change. In particular, the
exponent $\kappa$ of the logarithm of $t$ that appears in the
corresponding scaling variable is fully consistent with the droplet
prediction $\kappa=d/(d-1)=3/2$, with an accuracy of about 10\%. 
For the quench protocols considered here, our results again show
that the scaling behavior is consistent with the exponent $\kappa=3/2$
(in this case with an accuracy of approximately 20\%).

\section{Conclusions}
\label{conclu}

We investigate the out-of-equilibrium relaxational dynamics of 3D
$q$-state Potts models after varying the temperature across the FOT
point $\beta_{\rm fo}$.  In particular, we study the post-quench
dynamics arising from a sudden change of the inverse temperature
across the FOT, from the high-temperature to the low-temperature
phase, driven by a purely relaxational heat-bath dynamics.

We focus on the emergence of spinodal-like behaviors in the
thermodynamic limit, associated with the eventual dynamic phase
change.  We argue that, if the nucleation of droplets is the
relevant mechanism for the post-quench phase change, the relevant
scaling variable is $\rho = (\ln t)^{\kappa} \delta$ with
$\delta = \beta_f/\beta_{\rm fo}-1$ and $\kappa=d/(d-1)$ ($\kappa=3/2$
for $d=3$). The time-dependent infinite-volume energy density is expected
to be discontinuous for $\delta\to 0$ at fixed $\rho$ at some finite
value $\rho_s>0$.  The singularity in terms of the scaling
variable $\rho$ implies the emergence of a spinodal-like behavior,
with a time scale $\tau$ that increases exponentially as in
Eq.~(\ref{taubeh}), i.e., $\ln \tau \sim \delta^{-2/3}$ for decreasing
$\delta$.

The post-quench evolution of 3D $q$-state Potts models across their
thermal FOT is expected to behave as described above. This is confirmed
by a detailed analysis of the post-quench dynamics in the 3D $q=6$
Potts model. The same scaling behavior is expected for any
$q \geq 3$, the physical mechanism at the basis of the phase 
change being independent of $q$. This is also supported by earlier
results in two and three dimensions (see, e.g.,
Refs.~\cite{PV-17, PRV-26}, which discuss the evolution of the system
when the temperature is slowly varied across the first-transition point), 
which show that the scaling behavior is independent of $q$. Our present
results provide another example of
out-of-equilibrium behavior in which droplet nucleation is the
relevant mechanism controlling the time dependence of the phase
change. Similar behaviors emerge when considering (instantaneous and
slow) protocols across the magnetic low-temperature FOT of 2D Ising
systems~\cite{PV-26, PRV-26-3} and the thermal FOT of the 2D Potts
models. On the other hand, a different scaling behavior is observed at
the magnetic FOTs in 3D and 4D Ising systems~\cite{PV-26}, likely due
to different relevant mechanisms.

\end{document}